# What You Should Know About Megaprojects, and Why: An Overview

By

Bent Flyvbjerg

Professor and Founding Chair of Major Programme Management

Said Business School

Oxford University

Draft 9.2

# Abstract

This paper takes stock of megaproject management, an emerging and hugely costly field of study. First, it answers the question of how large megaprojects are by measuring them in the units mega, giga, and tera, concluding we are presently entering a new "tera era" of trillion-dollar projects. Second, total global megaproject spending is assessed, at USD 6-9 trillion annually, or 8 percent of total global GDP, which denotes the biggest investment boom in human history. Third, four "sublimes" – political, technological, economic, and aesthetic – are identified to explain the increased size and frequency of megaprojects. Fourth, the "iron law of megaprojects" is laid out and documented: Over budget, over time, over and over again. Moreover, the "break-fix model" of megaproject management is introduced as an explanation of the iron law. Fifth, Albert O. Hirschman's theory of the Hiding Hand is revisited and critiqued as unfounded and corrupting for megaproject thinking in both the academy and policy. Sixth, it is shown how megaprojects are systematically subject to "survival of the unfittest," explaining why the worst projects get built instead of the best. Finally, it is argued that the conventional way of managing megaprojects has reached a "tension point," where tradition is challenged and reform is emerging.

*Keywords*: Megaproject management; Scale; Four sublimes; Iron law of megaprojects; Break-fix model of megaprojects; Hirschman's Principle of the Hiding Hand; Survival of the unfittest; Tension points

# Mega, Giga, Tera: How Big Are Megaprojects?

Megaprojects are large-scale, complex ventures that typically cost a billion dollars or more, take many years to develop and build, involve multiple public and private stakeholders, are transformational, and impact millions of people.[1] Hirschman (1995: vii, xi) calls such projects "privileged particles of the development process" and points out that often they are "trait making," that is, they are designed to ambitiously change the structure of society, as opposed to smaller and more conventional projects that are "trait taking," i.e., they fit into pre-existing structures and do not attempt to modify these. Megaprojects, therefore, are not just magnified versions of smaller projects. Megaprojects are a completely different breed of project in terms of their level of aspiration, lead times, complexity, and stakeholder involvement. Consequently, they are also a very different type of project to manage. A colleague likes to say that if managers of conventional projects need the equivalent of a driver's license to do what they do then managers of megaprojects need a pilot's jumbo jet license.[2] And just like you would not want someone with only a driver's license to fly a jumbo, you don't want conventional project managers to manage megaprojects.

Megaprojects are increasingly used as the preferred delivery model for goods and services across a range of businesses and sectors, like infrastructure, water and energy, information technology, industrial processing plants, mining, supply chains, enterprise systems, strategic corporate initiatives and change programs, mergers and acquisitions, government administrative systems, banking, defense, intelligence, air and space exploration, big science, urban regeneration, and major events. Examples of megaprojects are high-speed rail lines, airports, seaports, motorways, hospitals, national health or pension ICT systems, national broadband, the Olympics, large-scale signature architecture, dams, wind farms, offshore oil and gas extraction, aluminum smelters, the development of new aircrafts, the largest container and cruise ships, high-energy particle accelerators, and the logistics systems used to run large supply-chain-based companies like Amazon and Maersk. Below we will see just how big megaprojects and the megaprojects business are. We will also understand what drives scale.

To illustrate just how big megaprojects are, consider one of the largest dollar figures from public economic debate in recent years, the size of US debt to China. This debt is around one trillion US dollars and is considered so large it may destabilize the world economy if the debt is not managed prudently. With this supersize measuring rod, now consider the fact that the combined cost of just two of the world's largest megaprojects – the Joint Strike Fighter aircraft program and China's high-speed rail project – is more than half of this figure, at 700 billion dollars (see Figure 1). The cost of a mere handful of the largest megaprojects in the world will dwarf almost any other economic figure, and certainly any investment figure.

[Figure 1 app. here]

However, not only are megaprojects large, they are constantly growing ever larger in a long historical trend with no end in sight. When New York's Chrysler Building opened in 1930 at 319 meters it was the tallest building in the world. The record has since been surpassed seven times and from 1998 the tallest building has significantly been located in emerging economies with Dubai's Burj Khalifa presently holding the record at 828 meters. That is a 160 percent increase in building height over 80 years. Similarly, the longest bridge span has grown even faster, by 260 percent over approximately the same period. Measured by value, the size of infrastructure projects has grown by 1.5 to 2.5 percent annually in real terms over the past century, which is equivalent to a doubling in project size two to three times per century (author's megaprojects database). The size of ICT projects, the new kid on the block, has grown much faster, as illustrated by a 16-fold increase from 1993 to 2009 in lines of code in Microsoft Windows, from five to 80 million lines. Other types of megaprojects, from the Olympics to industrial projects, have seen similar developments. Coping with increased scale is therefore a constant and pressing issue in megaproject management.

"Mega" comes from the Greek word "megas" and means great, large, vast, big, high, tall, mighty, and important. As a scientific and technical unit of measurement "mega" specifically means a million. If we were to use this unit of measurement in economic terms, then strictly speaking megaprojects would be million-dollar (or euro, pound, etc.) projects, and for more than a hundred years the largest projects in the world were indeed measured mostly in the millions. This changed with the Second World War, Cold War, and Space Race. Project costs now escalated to the billions, led by the Manhattan Project (1939-46), a research and development program that produced the first atomic bomb, and later the Apollo program (1961-72), which landed the first humans on the moon (Morris, 1994; Flyvbjerg, 2014). According to Merriam-Webster, the first known use of the term "megaproject" was in 1976, but before that, from 1968, "mega" was used in "megacity" and later, from 1982, as a standalone adjective to indicate "very large."

Thus the term "megaproject" caught on just as the largest projects technically were megaprojects no more, but, to be accurate, "gigaprojects" – "giga" being the unit of measurement meaning a billion. However, the term "gigaproject" never really caught on. A Google search reveals that the word "megaproject" is used 27 times more frequently on the web than the term "gigaproject". For the largest of this type of project, costs of 50-100 billion dollars are now common, as for the California and UK high-speed rail projects, and costs above 100 billion dollars not uncommon, as for the International

Space Station and the Joint Strike Fighter. If they were nations, projects of this size would rank among the world's top 100 countries measured by gross domestic product, larger than the economies of, for example, Kenya or Guatemala. When projects of this size go wrong, whole companies and national economies suffer.

"Tera" is the next unit up, as the measurement for a trillion (a thousand billion). Recent developments in the size of the very largest projects and programs indicate we may presently be entering the "tera era" of large-scale project management. If we consider as projects the stimulus packages that were launched by the United States, Europe, and China to mitigate the effects of the 2008 financial and economic crises, then we may speak of trillion-dollar projects and thus of "teraprojects." Similarly, if the major acquisition program portfolio of the United States Department of Defense – which was valued at 1.6 trillion dollars in 2013 – is considered a large-scale project, then this, again, would be a teraproject (United States Government Accountability Office, 2013: 2). Projects of this size compare with the GDP of the world's top 20 nations, similar in size to the national economies of for example Australia or Canada. There is no indication that the relentless drive to scale is abating in megaproject development. Quite the opposite; scale seems to be accelerating.

## How Big Is the Megaprojects Business?

But megaprojects are not only large and growing constantly larger, they are also being built in ever greater numbers at ever greater value. The McKinsey Global Institute (2013) estimates global infrastructure spending at USD 3.4 trillion per year 2013-2030, or approximately four percent of total global gross domestic product, mainly delivered as large-scale projects. *The Economist* (June 7, 2008: 80) similarly estimated infrastructure spending in emerging economies at USD 2.2 trillion annually for the period 2009-2018.

To illustrate the accelerated pace at which spending is taking place, consider that in the five years from 2004 to 2008, China spent more on infrastructure in real terms than in the whole of the 20[th] Century. That is an increase in spending rate of a factor twenty. Similarly, from 2005 to 2008, China built as many kilometers of high-speed rail as Europe did in two decades, and Europe was extraordinarily busy building this type of infrastructure during this period. Not at any time in the history of mankind has infrastructure spending been this high measured as a share of world GDP, according to *The Economist*, who calls it "the biggest investment boom in history." And that's just infrastructure.

If we include the many other fields where megaprojects are a main delivery model – oil and gas, mining, aerospace, defense, ICT, supply chains, mega events, etc. – then a conservative estimate for the global megaproject market is USD 6-9 trillion per year, or approximately eight percent of total global gross domestic product. For perspective, consider this is equivalent to spending five to eight times the accumulated US debt to China, *every year*. That's big business by any definition of the term.

Moreover, megaprojects have proved remarkably recession proof. In fact, the downturn from 2008 helped the megaprojects business grow further by showering stimulus spending on everything from transportation infrastructure to ICT. From being a fringe activity – albeit a spectacular one – mainly reserved for rich, developed nations, megaprojects have recently transformed into a global multi-trillion-dollar business that affects all aspects of our lives, from our electricity bill to how we shop and what we do on the Internet to how we commute.

With so many resources tied up in ever-larger and ever-more megaprojects, at no time has the management of such projects been more important. The potential benefits of building the right projects in the right manner are enormous and are only matched by the potential waste from building the wrong projects, or building projects wrongly. Never has it been more important to choose the most fitting projects and get their economic, social, and environmental impacts right (Flyvbjerg el al., 2003). Never has systematic and valid knowledge about megaprojects therefore been more important to inform policy, practice, and public debate in this highly costly area of business and government.

## The Four Sublimes

What drives the megaproject boom described above? Why are megaprojects so attractive to decision makers? The answer may be found in the so-called "four sublimes" of megaproject management (see Table 1). The first of these, the "technological sublime," is a term variously attributed to Miller (1965) and Marx (1967) to describe the positive historical reception of technology in American culture during the nineteenth and early twentieth centuries. Frick (2008) introduced the term to the study of megaprojects and here describes the technological sublime as the rapture engineers and technologists get from building large and innovative projects with their rich opportunities for pushing the boundaries for what technology can do, like building the tallest building, the longest bridge, the fastest aircraft, the largest wind turbine, or the first of anything. Frick applied the concept in a case study of the multi-billion-dollar New San Francisco-Oakland Bay Bridge, concluding "the technological sublime dramatically influenced bridge design, project outcomes, public debate, and the lack of accountability for its [the bridge's] excessive cost overruns" (239).

Flyvbjerg (2012, 2014) proposed three additional sublimes, beginning with the "political sublime," which here is understood as the rapture politicians get from building monuments to themselves and their causes. Megaprojects are manifest, garner attention, and lend an air of proactiveness to their promoters. Moreover, they are media magnets, which appeals to politicians who seem to enjoy few things better than the visibility they get from starting megaprojects. Except maybe cutting the ribbon of one in the company of royals or presidents, who are likely to be present lured by the unique monumentality and historical import of many megaprojects. This is the type of public exposure that helps get politicians re-elected. They therefore actively seek it out.

Next there is the "economic sublime," which is the delight business people and trade unions get from making lots of money and jobs off megaprojects. Given the enormous budgets for megaprojects there are ample funds to go around for all, including contractors, engineers, architects, consultants, construction and transportation workers, bankers, investors, landowners, lawyers, and developers. Finally, the "aesthetic sublime" is the pleasure designers and people who appreciate good design get from building, using, and looking at something very large that is also iconically beautiful, like San Francisco's Golden Gate bridge or Sydney's Opera House.

All four sublimes are important drivers of the scale and frequency of megaprojects described above. Taken together they ensure that strong coalitions exist of stakeholders who benefit from megaprojects and who will therefore work for more such projects.

[Table 1 app. here]

For policy makers, investment in infrastructure megaprojects seems particularly coveted, because, if done right, such investment:

- Creates and sustains employment.
- Contains a large element of domestic inputs relative to imports.
- Improves productivity and competitiveness by lowering producer costs.
- Benefits consumers through higher-quality services.
- Improves the environment when infrastructures that are environmentally sound replace infrastructures that are not (Helm, 2008: 1).

But there is a big "if" here, as in "if done right." Only if this is disregarded – as it often is by promoters and decision makers for megaprojects – can megaprojects be seen as an effective way to deliver infrastructure. In fact, conventional megaproject delivery – infrastructure and other – is highly problematic with a dismal performance record in terms of actual costs and benefits, as we will see below. The following characteristics of megaprojects are typically overlooked or glossed over when the four sublimes are at play and the megaproject format is chosen for delivery of large-scale ventures:

1. Megaprojects are inherently risky due to long planning horizons and complex interfaces (Flyvbjerg, 2006).

2. Often projects are led by planners and managers without deep domain experience who keep changing throughout the long project cycles that apply to megaprojects, leaving leadership weak.

3. Decision-making, planning, and management are typically multi-actor processes involving multiple stakeholders, public and private, with conflicting interests (Aaltonen and Kujala, 2010).

4. Technology and designs are often non-standard, leading to "uniqueness bias" amongst planners and managers, who tend to see their projects as singular, which impedes learning from other projects.[3]

5. Frequently there is overcommitment to a certain project concept at an early stage, resulting in "lock-in" or "capture," leaving alternatives analysis weak or absent, and leading to escalated commitment in later stages. "Fail fast" does not apply; "fail slow" does (Cantarelli et al., 2010; Ross and Staw, 1993; Drummond, 1998).

6. Due to the large sums of money involved, principal-agent problems and rent-seeking behavior are common, as is optimism bias (Eisenhardt, 1989; Stiglitz, 1989; Flyvbjerg el al., 2009).

7. The project scope or ambition level will typically change significantly over time.

8. Delivery is a high-risk, stochastic activity, with overexposure to so-called "black swans," i.e., extreme events with massively negative outcomes (Taleb, 2010). Managers tend to ignore this, treating projects as if they exist largely in a deterministic Newtonian world of cause, effect, and control.

9. Statistical evidence shows that such complexity and unplanned events are often unaccounted for, leaving budget and time contingencies inadequate.

10. As a consequence, misinformation about costs, schedules, benefits, and risks is the norm throughout project development and decision-making. The result is cost overruns, delays, and benefit shortfalls that undermine project viability during project implementation and operations.

In the next section, we will see just how big and frequent such cost overruns, delays, and benefit shortfalls are.

## The Iron Law of Megaprojects

Performance data for megaprojects speak their own language. Nine out of ten such projects have cost overruns. Overruns of up to 50 percent in real terms are common, over 50 percent not uncommon. Cost overrun for the Channel tunnel, the longest underwater rail tunnel in Europe, connecting the UK and France, was 80 percent in real terms. For Denver International Airport, 200 percent. Boston's Big Dig, 220 percent. The UK National Health Service IT system, 400-700 percent. The Sydney Opera House, 1,400 percent (see more examples in Table 2). Overrun is a problem in private as well as public sector projects, and things are not improving; overruns have stayed high and constant for the 70-year period for which comparable data exist. Geography also does not seem to matter; all countries and continents for which data are available suffer from overrun. Similarly, benefit shortfalls of up to 50 percent are also common, and above 50 percent not uncommon, again with no signs of improvements over time and geography (Flyvbjerg et al., 2002, 2005).

[Table 2 app. here]

Combine the large cost overruns and benefit shortfalls with the fact that business cases, cost-benefit analyses, and social and environmental impact assessments are typically at the core of planning and decision-making for megaprojects and we see that such analyses can generally not be trusted. For instance, for rail projects an average cost overrun of 44.7 percent combines with an average demand shortfall of 51.4 percent, and for roads, an average cost overrun of 20.4 percent combines with a fifty–fifty risk that demand is also wrong by more than 20 percent. With errors and biases of such magnitude in the forecasts that form basis for business cases, cost–benefit analyses, and social and environmental impact assessments, such analyses will also, with a high degree of certainty, be strongly misleading. "Garbage in, garbage out," as the saying goes (Flyvbjerg, 2009).

As a case in point, consider the Channel tunnel in more detail. This project was originally promoted as highly beneficial both economically and financially. At the initial public offering, Eurotunnel, the private owner of the tunnel, tempted investors by telling them that 10 percent "would be a reasonable allowance for the possible impact of unforeseen circumstances on construction costs."[4] In fact, costs went 80 percent over budget for construction, as mentioned above, and 140 percent for financing.

Revenues have been half of those forecasted. As a consequence the project has proved non-viable, with an internal rate of return on the investment that is negative, at minus 14.5 percent with a total loss to the British economy of 17.8 billion US dollars. Thus the Channel tunnel detracts from the economy instead of adding to it. This is difficult to believe when you use the service, which is fast, convenient, and competitive with alternative modes of travel. But in fact each passenger is heavily subsidized. Not by the taxpayer this time, but by the many private investors who lost their money when Eurotunnel went insolvent and was financially restructured. This drives home an important point: A megaproject may well be a technological success, but a financial failure, and many are. An economic and financial ex post evaluation of the Channel tunnel, which systematically compared actual with forecasted costs and benefits, concluded that "the British Economy would have been better off had the Tunnel never been constructed" (Anguera, 2006: 291). Other examples of non-viable megaprojects are Sydney's Lane Cove tunnel, the high-speed rail connections at Stockholm and Oslo airports, the Copenhagen metro, and Denmark's Great Belt tunnel, the second-longest under-water rail tunnel in Europe, after the Channel tunnel.

Large-scale ICT projects are even more risky. One in six such projects become a statistical outlier in terms of cost overrun with an average overrun for outliers of 200 percent in real terms. This is a 2,000 percent overincidence of outliers compared to normal and a 200 percent overincidence compared to large construction projects, which are also plagued by cost outliers (Flyvbjerg and Budzier, 2011). Total project waste from failed and underperforming ICT projects for the United States alone has been estimated at 55 billion dollars annually by the Standish Group (2009).

Delays are a separate problem for megaprojects and delays cause both cost overruns and benefit shortfalls. For instance, preliminary results from a study undertaken at Oxford University, based on the largest database of its kind, suggest that delays on dams are 45 percent on average. Thus if a dam was planned to take 10 years to execute, from the decision to build until the dam became operational, then it actually took 14.5 years on average. Flyvbjerg et al. (2004) modeled the relationship between cost overrun and length of implementation phase based on a large data set for major construction projects. They found that on average a one-year delay or other extension of the implementation phase correlates with an increase in percentage cost overrun of 4.64 percent.

To illustrate, for a project the size of London's 26 billion dollars Crossrail project, a one-year delay would cost 1.2 billion dollars extra, or 3.3 million dollars per day. The key lesson here is that in order to keep costs down, implementation phases should be kept short and delays small. This should not be seen as an excuse for fast-tracking projects, i.e., rushing them through decision making for early construction start. Front-end planning needs to be thorough before deciding whether to give the green light to a project or stopping it (Williams and Samset, 2010). But often the situation is the exact opposite. Front-end planning is scant, bad projects are not stopped, implementation phases and delays are long, costs soar, and benefits and revenue realization recedes into the future. For debt-financed projects this is a recipe for disaster, because project debt grows while there is no revenue stream to service interest payments, which are then added to the debt, etc. As a result, many projects end up in the so-called "debt trap" where a combination of escalating construction costs, delays, and increasing interest payments makes it impossible for income from a project to cover costs, rendering the project non-viable. That is what happened to the Channel tunnel and Sydney's Lane Cove tunnel, among other projects.

This is not to say projects do not exist that were built on budget and time and delivered the promised benefits. The Guggenheim Museum Bilbao is an example of that rare breed of project. Similarly, recent metro extensions in Madrid were built on time and to budget (Flyvbjerg, 2005) as were a number of industrial projects (Merrow, 2011). It is particularly important to study such projects to understand the causes of success and test whether success may be replicated elsewhere. It is far easier, however, to produce long lists of projects that have failed in terms of cost overruns and benefit shortfalls than it is to produce lists of projects that have succeeded. To illustrate, as part of ongoing research on success in megaproject management the present author and his associates are trying to establish a sample of successful projects large enough to allow statistically valid answers. But so far they have failed. Why? Because success is so rare in megaproject management that at present it can be studied only as small-sample research, whereas failure may be studied with large samples of projects.

Success in megaproject management is typically defined as projects being delivered on budget, time, and benefits. If, as the evidence indicates, approximately one out of ten megaprojects is on budget, one out of ten is on schedule, and one out of ten is on benefits, then approximately one in a thousand projects is a success, defined as on target for all three. Even if the numbers were wrong by a factor two – so that two, instead of one, out of ten projects were on target for cost, schedule, and benefits, respectively – the success rate would still be dismal, now eight in a thousand. This serves to illustrate what may be called the "iron law of megaprojects": *Over budget, over time, over and over again* (Flyvbjerg, 2011).[5] Best practice is an outlier, average practice a disaster in this interesting and very costly area of management.

## The "Break-Fix Model" of Megaproject Management

The above analysis leaves us with a genuine paradox, the so-called "megaprojects paradox," first identified by Flyvbjerg et al. (2003: 1-10). On one side of the paradox, megaprojects as a delivery model for public and private ventures have never been more in demand, and the size and frequency of megaprojects have never been larger. On the other side, performance in megaproject management is strikingly poor and has not improved for the 70-year period for which comparable data are available, at least not when measured in terms of cost overruns, schedule delays, and benefit shortfalls.

Today, megaproject planners and managers are stuck in this paradox because their main delivery method is what has been called the "break-fix model" for megaproject management.[6] Generally, megaproject planners and managers – and their organizations – do not know how to deliver successful megaprojects, or do not have the incentives to do so, and therefore such projects tend to "break" sooner or later, for instance when reality catches up with optimistic, or manipulated, estimates of schedule, costs, or benefits; and delays, cost overruns, etc. follow. Projects are then often paused and reorganized – sometimes also refinanced – in an attempt to "fix" problems and deliver some version of the initially planned project with a semblance of success. Typically lock-in and escalation make it impossible to drop projects altogether, which is why megaprojects have been called the "Vietnams" of policy and management: "easy to begin and difficult and expensive to stop" (White, 2012; also Cantarelli el al., 2010; Ross and Staw, 1993, Drummond, 1998). The "fix" often takes place at great and unexpected cost to those stakeholders who were not in the know of what was going on and were unable to or lacked the foresight to pull out before the break.

The break-fix model is wasteful and leads to misallocation of resources, in both organizations and society, for the simple reason that under this model decisions to go ahead with projects are based on misinformation more than on information. The degree of misinformation varies significantly from project to project, as documented by the large standard deviations that apply to cost overruns and benefit shortfalls (Flyvbjerg et al., 2002, 2005). We may therefore *not* assume, as is often done, that on average all projects are misrepresented by approximately the same degree and, therefore, we are still building the best projects, even if they are not as good as they appear on paper. The truth is, we don't know, and often projects turn out to bring a net loss to the economy, instead of a gain. The cure to the break-fix model is to get projects right from the outset so they don't break, through proper front-end management.

## Hirschman's Hiding Hand, Revisited

One may argue, of course, as famously done by Hirschman (1967a: 12-13), that if people knew in advance the real costs and challenges involved in delivering a large project, "they probably would never have touched it" and nothing would ever get built. So it is better not to know, because ignorance helps get projects started, according to this argument. The following is a recent and particularly candid articulation of the nothing-would-ever-get-built argument, by former California State Assembly speaker and mayor of San Francisco, Willie Brown, discussing a large cost overrun on the San Francisco Transbay Terminal megaproject in his *San Francisco Chronicle* column (July 28, 2013, emphasis added):

> "News that the Transbay Terminal is something like $300 million over budget should not come as a shock to anyone. We always knew the initial estimate was way under the real cost. Just like we never had a real cost for the [San Francisco] Central Subway or the [San Francisco-Oakland] Bay Bridge or any other massive construction project. So get off it. In the world of civic projects, the first budget is really just a down payment. *If people knew the real cost from the start, nothing would ever be approved.* The idea is to get going. Start digging a hole and make it so big, there's no alternative to coming up with the money to fill it in."

Rarely has the tactical use by project advocates of cost underestimation, sunk costs, and lock-in to get projects started been expressed by an insider more plainly, if somewhat cynically. It is easy to obtain such statements off the record, but few are willing to officially lend their name to them, for legal and ethical reasons to which we will return. Nevertheless, the nothing-would-ever-get-built argument has been influential with both practitioners and academics in megaproject management. The argument is deeply flawed, however, and thus deserves a degree of attention and critique. Hirschman's text contains the classic formulation of the argument and has served widely as its theoretical justification, as has Sawyer (1952), who directly inspired and influenced Hirschman.[7] A recent celebration of Hirschman's thinking on this point may be found in Gladwell (2013).

Hirschman (1967a: 13-14) observed that humans are "tricked" into doing big projects by their own ignorance. He saw this as positive because just as humans underestimate the difficulties in doing large-scale projects they also underestimate their own creativity in dealing with the difficulties, he believed, and "the only way in which we can bring our creative sources fully into play is by misjudging the nature of the task, by presenting it to ourselves as more routine, simple, undemanding of genuine creativity than it will turn out to be." Hirschman called this the "principle of the Hiding Hand" and it consists of "some sort of invisible or hidden hand that beneficially hides difficulties for us," where the error of underestimating difficulties is offset by a "roughly similar" error in underestimating our ability to overcome the difficulties thus helping "accelerate the rate at which 'mankind' engages successfully in problem-solving."

Sawyer (1952: 199, 203) – in a study of early industrial infrastructure projects that he called a work "in praise of folly" – similarly identified what he called "creative error" in project development as, first, "miscalculation or sheer ignorance" of the true costs and benefits of projects and, second, such miscalculation being "crucial to getting an enterprise launched at all." Sawyer argued that such "creative error" was key to building a number of large and historically important projects like the Welland Canal between Lake Erie and Lake Ontario, the Panama Canal, the Middlesex Canal, the Troy and Greenfield Railroad, and early Ohio roads. For these and other projects, Sawyer found that "the error in estimating costs was at least offset by a corresponding error in the estimation of demand" (p. 200). Hirschman (1967a: 16) explicitly mentioned Sawyer as an inspiration and his "creative error" as a close "approximation" to the Hiding Hand principle.

It is easy to understand why Hirschman's and Sawyer's theories have become popular, especially with people who benefit from megaprojects. The theories encourage promoters and decision makers, like Willie Brown above, to just go ahead with projects and not worry too much about the costs or other problems, because the Hiding Hand will take care of them, eventually. And, in any case, who wants to be the killjoy stopping large projects from going ahead by an overdose of truth? Hirschman (1967b) was an immediate hit with practitioners, from Washington's policy establishment to the United Nations to the World Bank. The head of the bank's Economics Department told Hirschman, "You've helped in part to remove the unease that I have had in reflecting on the fact that if our modern project techniques had been used, much of the existing development in the world would never have been undertaken" (Adelman, 2013: 405). Hirschman's thinking also eventually penetrated academia. Teitz and Skaburskis (2003) follow the Hiding Hand logic when they ask of the huge cost overrun on the Sydney Opera House, "Did people really think that the Sydney Opera House would come in on budget? Or did we all agree to accept the deception and engage in wishful thinking in order to make something that we really wanted happen? ... [D]o Australians really regret those dramatic sails in the harbour? Or would they have regretted more the decision [not to build] that would most reasonably have been based on a fair prediction of costs?"

The logic is seductive, yet precarious. In retrospect, of course Australians do not regret the Sydney Opera House, given what it has done for Australia – though at first the building was not called "dramatic sails in the harbour," but "copulating white turtles" and "something that is crawling out of the ocean with nothing good in mind" designed by an architect with "lousy taste" (Reichold and Graf, 2004: 168). Non-Australians may feel regret, however, for instance the architect of the Opera House, what's his name? Does anybody know? Only few do, which seems surprising given we are talking about the architect of arguably the most iconic building of the 20th century. And if anybody knows the architect is the Dane Jørn Utzon, how come they can hardly ever mention another building designed by him? Because the overrun on the Opera House and the following controversy destroyed Utzon's career and kept him from building more masterpieces. He became that most tragic figure in architecture, the one-building-architect. This is the real regret – and real cost – of the Sydney Opera House. Not premier Joe Cahill's deliberate deception about the cost – to get approval in Parliament – and the consequential huge cost overrun (Flyvbjerg, 2005).

In a meeting held in support of Utzon at Sydney Town Hall in March 1966 – six weeks before the controversy made Utzon leave Australia and the Opera House, in the middle of construction and never to return – the Vienneseborn Australian architect Harry Seidler said, "If Mr. Utzon leaves, a crime will have been committed against future generations of Australians" (Murray, 2004: 105). Seidler was more right than he could have imagined, except the crime would not be limited to Australians, it became a crime against lovers of great architecture everywhere. After winning the Pritzker Prize – the Nobel of architecture – in 2003, Utzon again became widely acclaimed, even in Australia, where the Sydney Opera tour guides for years had been forbidden to even mention his name. But it was too late. Utzon was now 85 and he had not built anything major for decades. So instead of having a whole oeuvre to enjoy, as we have for other architects of his caliber, we have just the one main building. Utzon was 38 when he won the competition for the Opera House – how would the work of the mature master have enriched our lives? We will never know.

As a thought experiment, consider the collected works of architect Frank Gehry, who is in the same league as Utzon; then consider which building you would choose, could you choose only one, and the rest would have to go. So if you chose, say, the Guggenheim Museum Bilbao, then Los Angeles' Disney Concert Hall, Chicago's Jay Pritzker Pavilion, Prague's Dancing House, Seattle's Experience Music Project Museum, etc. would be eliminated. This illustrates the high price the government of New South Wales has imposed on the world by mismanaging the planning of the Sydney Opera House and deliberately playing the game of creative error and Hiding Hand. Even if the Opera House may be an extreme case, Sydney drives home an important point: managing by creative error is risky and disruptive, sometimes in drastic and unexpected ways, and the Hiding Hand isn't big enough to hide all, or even most, errors.

Hirschman's and Sawyer's theories are also flawed at a more basic level, that of validity. A close look reveals the theories to be based on small samples and biased data. Hirschman studied only 11 projects, or a few more if we count subprojects, Sawyer ten to 15. This important fact is typically ignored when the Hiding Hand principle is discussed. Hirschman (1967a: 7, 14) seemed aware of the weak foundations and limited applicability of the principle when he called it "speculative" and useful only "[u]p to a point." To a colleague he admitted at the time of publication that his book was "an exploration, an experiment;" to another he said he had deliberately biased his analysis "to emphasize unexpected successes" (Adelman, 2013: 404-5). Even so, Hirschman went on to call the Hiding Hand a "general principle of action" and brazenly used a name for it with clear connotations to Adam Smith's famous Invisible (Hidden) Hand. Evidently, the temptation to formulate an "economic law" was too strong, despite the weak and biased data. Sawyer (1952: 204) warned the reader up front that his study must be considered a "marginal and distinctly limited note." He admitted the study considers only a "quite special kind of case" and neglects projects that were "failures" in order to focus on projects that were "successful" in the sense that "an original gross miscalculation as to costs ... was happily offset by at least a corresponding underestimation of demand." Sawyer's results thus do not describe a general characteristic of large projects, but a characteristic of his biased sample that includes only projects lucky enough to have had large underestimates of costs compensated by similarly large or larger underestimates of demand. Some would call this data fishing and the only redeeming factor is that Sawyer was disarmingly honest and tongue-in-cheek humoristic about it. He appears to not have expected to be taken wholly seriously, which unfortunately he was by some, including Hirschman.

Today we have much better data and theories on megaproject performance than at the time of Hirschman and Sawyer. We now know that, while there may be elements of truth in these authors' theories for certain types of projects and contexts, their samples and conclusions are not representative of the project population. In particular, their odd asymmetrical assumption that optimism would apply to cost estimates but pessimism to estimates of benefits has been solidly disproved by Kahneman and Tversky (1979a, b) and behavioral economists building on their work. They found that optimism bias applies to estimates of costs and benefits, both. An optimistic cost estimate is low and leads to cost overrun, whereas an optimistic benefit estimate is high and results in benefit shortfalls. Thus errors of estimation do not cancel each other out, as Hirschman would have it; the exact opposite happens, errors generally reinforce each other.

Megaproject planners and managers would therefore be ill advised to count on Hiding Hands, creative errors, or any other general principle according to which underestimates of costs would be balanced by similar underestimates of benefits. We also now know it would be equally foolhardy to assume that downstream human creativity may be generally counted on to solve problems that planners and managers overlook or underestimate when the decision is made to go ahead with a project. The data show that for too many projects with front-end problems such creativity never materializes and projects end up seriously impaired or non-viable. Initial problems, if not dealt with up front, tend not to go away. The iron law of megaprojects, described above, trumps Hirschman's Hiding Hand at a high level of statistical significance, and we know why. The Hiding Hand is itself an example of optimism and does therefore not capture the reality of megaproject management. For such capture, and true explanatory power, we must turn to theories of optimism bias, the planning fallacy, strategic misrepresentation, and principal-agent behavior.

## Survival of the Unfittest

In sum, one does megaprojects – and megaproject management – a disservice if one claims they can only be done through the Hiding Hand, creative error, or downright deception. It is, undoubtedly, quite common for project promoters and their planners and managers to believe their projects will benefit society and that, therefore, they are justified in "cooking" costs and benefits to get projects built (Wachs, 1990; Pickrell, 1992). Such reasoning is faulty, however. Underestimating costs and overestimating benefits for a given project – which is the common pattern, as described above – leads to a falsely high benefit-cost ratio for that project, which in turn leads to two problems. First, the project may be started despite the fact it is not financially and economically viable. Or, second, it may be started instead of another project that would have shown itself to yield higher returns than the project started, had the real costs and benefits of both projects been known. Both cases result in Pareto inefficiency, that is, the misallocation of resources and, for public projects, waste of taxpayers' money. Thus for reasons of economic efficiency alone the argument must be rejected that cost underestimation and benefit overestimation are justified to get projects started.

But the argument must also be rejected for legal and ethical reasons. In most democracies, for project promoters, planners, and managers to deliberately misinform legislators, administrators, bankers, the public, and the media about costs and benefits would not only be considered unethical but in some instances also unlawful, for instance where civil servants would intentionally misinform cabinet members, or cabinet members would intentionally misinform parliament. In private corporations, Sarbanes-Oxley-like legislation similarly makes deliberate misrepresentation a crime under many circumstances, which in the US is punishable with prison up to 20 years.[8] There is a formal "obligation to truth" built into most democratic constitutions – and now also in legislation for corporate

governance – as a means for enforcing accountability. This obligation would be violated by deliberate misrepresentation of costs and benefits, whatever the reasons for such misrepresentation may be. Not only economic efficiency would suffer but also democracy, good governance, and accountability.

A first answer to the skeptics' question of whether enough megaprojects would be undertaken if some form of misrepresentation of costs and benefits was not involved is, therefore, that even if misrepresentation was necessary in order to get projects started, such misrepresentation would typically not be defensible in liberal democracies – and especially not if it was deliberate – for economic, legal, and ethical reasons.

A second answer is that misrepresentation is not necessary to undertake projects, because many projects exist with sufficiently high benefits and low enough costs to justify building them. Even in the field of innovative and complex architecture, which is often singled out as particularly difficult, there is the Basque Abandoibarra urban regeneration project, including the Guggenheim Museum Bilbao, which is as complex, innovative, and iconic as any signature architecture, and was built on time and budget. Complex rail projects, too, like the Paris-Lyon high-speed rail line and the London Docklands light railway extension have been built to budget. The problem is not that projects worth undertaking do not exist or cannot be built on time and budget. The problem is that the dubious and widespread practices of underestimating costs and overestimating benefits used by many megaproject promoters, planners, and managers to promote *their* pet project create a distorted hall-of-mirrors in which it is extremely difficult to decide which projects deserve undertaking and which not.

In fact the situation is even worse than that. The common practice of depending on the Hiding Hand or creative error in estimating costs and benefits – thus "showing the project at its best" as an interviewee put it in a previous study – results in an inverted Darwinism, i.e., the "survival of the unfittest" (Flyvbjerg, 2009: 352). It is not the best projects that get implemented in this manner, but the projects that look best on paper. And the projects that look best on paper are the projects with the largest cost underestimates and benefit overestimates, other things being equal. But the larger the cost underestimate on paper, the greater the cost overrun in practice. And the larger the overestimate of benefits, the greater the benefit shortfall. Therefore the projects that have been made to look best on paper become the worst, or unfittest, projects in reality, in the sense that they are the very projects that will encounter most problems during construction and operations in terms of the largest cost overruns, benefit shortfalls, and risks of non-viability. They have been designed like that, as disasters waiting to happen.

The result is, as even the industry's own organ, the Major Projects Association, has said, that "too many projects proceed that should not have done" (Morris and Hough, 1987: 214). One might add that projects also exist that do not proceed but should have, had they not lost out, not to better projects but to projects with "better" creative error, that is "better" manipulated estimates of costs and benefits.

## Light at the End of the Tunnel?

Fortunately, signs of improvement in megaproject management have recently appeared. The tacit consensus that misrepresentation is an acceptable business model for project development is under attack. Shortly after taking office, President Obama openly identified "the costly overruns, the fraud and abuse, the endless excuses" in public procurement for major projects as key policy problems (White House, 2009). The *Washington Post* rightly called this "a dramatic new form of discourse" (Froomkin, 2009). Other countries are seeing similar developments. Before Obama it was not common in government or business to talk openly about overruns, fraud, and abuse in relation to megaprojects, although they were widespread then as now. The few who did so were ostracized. However, as emphasized by Wittgenstein (2009), we cannot solve problems we cannot talk about. So talking is the first step.

A more material driver of improvement is the fact that the largest projects are now so big and consequential in relation to individual businesses and agencies that cost overruns, benefit shortfalls, and risks from even a single project may bring down executives and whole corporations. This happened with the Airbus A380 superjumbo, when delays, cost overruns, and revenue shortfalls cost the CEO and other top managers their jobs. The CEO of BP was similarly forced to step down and the company lost more than half its value when the Deepwater Horizon offshore oil drilling rig caught fire and caused the world's largest oil spill in the Gulf of Mexico in 2010. At Kmart, a large US retailer, the entire company went bankrupt when a new multi-billion-dollar ICT enterprise system, which was supposed to make Kmart competitive with Walmart and Target, went off the rails (Flyvbjerg and Budzier, 2011). In China, corruption and related safety issues on the country's 300 billion dollar high-speed rail program have caused massive reputational damage, and cost the railway minister his political life in 2011. Today, if you are a CEO, minister, permanent secretary, or other top manager and want to be sure to keep your job, you will want to manage your megaprojects properly. Episodes like these have triggered leaders to begin looking for better megaproject delivery.

Even the wealth of whole cities and nations may be affected by a single megaproject failure. In Hong Kong, months of hiccups at the opening of a new international airport made traffic go elsewhere resulting in a fall in GNP for the entire city state. For Greece, a contributing factor to the country's 2011 debt default was the 2004 Athens Olympics, where cost overruns and incurred debt were so large they negatively affected the credit rating of the whole nation, substantially weakening the economy in the years before the 2008 international financial crisis. This resulted in a double dip, and disaster, for Greece, when other nations had only a single dip. Likewise, in Japan 2011, the nuclear tragedy at Fukushima significantly and negatively impacted the national economy as a whole. It is becoming increasingly clear that when megaprojects go wrong they are like the proverbial bull in the china shop: it takes just one to smash up the entire store. It is becoming similarly clear to many involved that something needs to be done about his.

In the UK at the beginning of the century, cost underestimation and overrun was rampant in so many projects in so many ministries that the reliability of national budgets suffered, leading the chancellor to order a Green Book on the problem and how to solve it (HM Treasury, 2003). This move inspired other countries to follow suit. Lawmakers and governments have begun to see that national fiscal distress and unreliable national budgets are too high a price to pay for the conventional way of managing megaprojects. In 2011, the UK Cabinet Office and HM Treasury joined forces to establish a Major Projects Authority with an enforceable mandate directly from the Prime Minister to oversee and direct the effective management of all large-scale projects that are funded and delivered by central government. In 2012, the Authority established, in collaboration with Oxford University, a Major Projects Leadership Academy – the first of its kind in the world – to train and authorize all UK civil servants in charge of central government major projects.[9]

Outside of government, private finance in megaprojects has been on the rise over the past twenty years. This means that capital funds, pension funds, and banks are increasingly gaining a say in management. Private capital is no panacea for the ills in megaproject management, to be sure; in some cases private capital may even make things worse (Hodge and Greve, 2009). But private investors place their own funds at risk. Funds and banks can therefore be observed to not automatically accept at face value the cost and revenue forecasts of project managers and promoters. Banks typically bring in their own advisers to do independent forecasts, due diligence, and risk assessments, which is an important step in the right direction (Flyvbjerg, 2013). The false assumption that one forecast or one business case may contain the whole truth about a project is problematized. Instead project managers and promoters are getting used to the healthy fact that different stakeholders hold different forecasts and that forecasts are not only products of data and mathematical modeling but also of power and negotiation. Why is this more healthy? Because it undermines trust in the misleading forecasts often produced by project promoters.

Moreover, democratic governance is generally getting stronger around the world. Corporate scandals, from Enron and onwards, have triggered new legislation and a war on corporate deception that is spilling over into government with the same objective: to curb waste and promote good governance. Although progress is slow, good governance is gaining a foothold even in megaproject management. The main drivers of reform come from outside the agencies and industries conventionally involved in megaprojects, which is good because it increases the likelihood of success. For example, the UK Treasury now requires that all ministries develop and implement procedures for megaprojects that will curb so-called "optimism bias" (Flyvbjerg, 2006). Funding will be unavailable for projects that do not take into account such bias, and methods have been developed for doing this (UK Department for Transport, 2006). Switzerland and Denmark have followed the lead of the UK (Swiss Association of Road and Transportation Experts, 2006; Danish Ministry for Transport and Energy, 2006, 2008). In Australia, the Parliament of Victoria has conducted an inquiry into how government may arrive at more successful delivery of significant infrastructure projects (Parliament of Victoria, 2012). Similarly, in the Netherlands the Parliamentary Committee on Infrastructure Projects did extensive public hearings to identify measures that will limit the misinformation about large infrastructure projects presented to the Parliament, public, and media (Dutch Commission on Infrastructure Projects, 2004). In Boston, the government has sued to recoup funds from contractor overcharges for the Big Dig related to cost overruns. More countries and cities are likely to follow the lead of the UK, Australia, Switzerland, Denmark, the Netherlands, and Boston in coming years.

Finally, research on how to reform megaproject management – examples of which has been referenced above – is beginning to positively impact practice. Such research has recently made great strides in better understanding what causes the many failures in megaproject delivery, and how to avoid them. For instance, we now understand that optimism bias and strategic misrepresentation are significantly better explanations of megaproject outcomes than previous explanations, including Hirschman's Hiding Hand and Sawyers creative error discussed above. And with a better understanding of causes has followed a better grasp of cures, from front-end management (Williams and Samset, 2010) to reference class forecasting (Kahneman, 2011: 243-254; Flyvbjerg, 2006) to institutional design for better accountability (Scott, 2012; Bruzelius et al., 1998). Moreover, research is beginning to help us understand success and how to replicate it. Perhaps most importantly, researchers have begun to take seriously the task of feeding their research results into the public sphere so they may effectively form part of public deliberation, policy, and practice (Flyvbjerg, 2012; Flyvbjerg et al., 2012).

With these developments things are moving in the right direction for megaproject management. It is too early to tell whether the reform measures being implemented will ultimately be successful. It seems unlikely, however, that the forces that have triggered the measures will be reversed, and it is those forces that reform-minded individuals and groups need to support and work with in order to improve megaproject management. This is the "tension point" where convention meets reform, power balances change, and new things are happening. In short, it is the place to be as a megaproject planner, manager, scholar, student, owner, or interested citizen.[10]

Figure 1: Size of selected megaprojects, measured against one of the largest dollar-figures in the world, accumulated US debt to China.

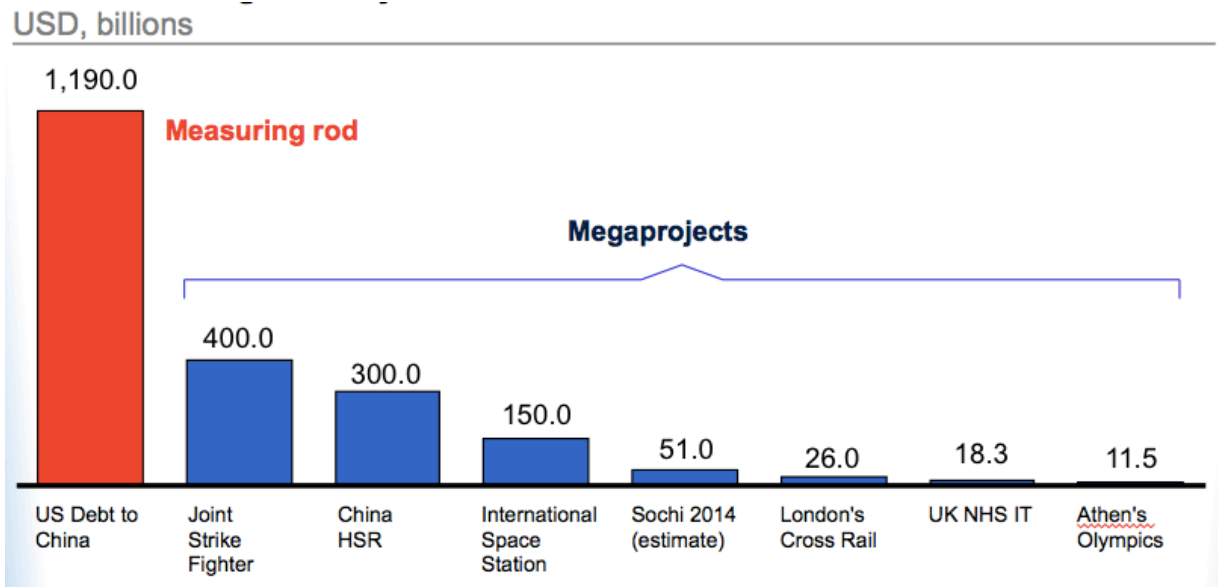

Table 1: The "Four Sublimes" that drive megaproject development.

| Type of Sublime | Characteristic |
|---|---|
| Political | The rapture politicians get from building monuments to themselves and their causes, and from the visibility this generates with the public and media |
| Technological | The excitement engineers and technologists get in pushing the envelope for what is possible in "longest-tallest-fastest" type of projects |
| Economic | The delight business people and trade unions get from making lots of money and jobs off megaprojects, including for contractors, workers in construction and transportation, consultants, bankers, investors, landowners, lawyers, and developers |
| Aesthetic | The pleasure designers and people who love good design get from building and using something very large that is also iconic and beautiful, like the Golden Gate bridge |

Table 2: Large-scale projects have a calamitous history of cost overrun.

| Project | Cost Overrun (%) |
|---|---|
| Suez Canal, Egypt | 1,900 |
| Scottish Parliament Building, Scotland | 1,600 |
| Sydney Opera House, Australia | 1,400 |
| Montreal Summer Olympics, Canada | 1,300 |
| Concorde supersonic aeroplane, UK, France | 1,100 |
| Troy and Greenfield railroad, USA | 900 |
| Excalibur Smart Projectile, USA, Sweden | 650 |
| Canadian Firearms Registry, Canada | 590 |
| Lake Placid Winter Olympics, USA | 560 |
| Medicare transaction system, USA | 560 |
| National Health Service IT system, UK | 550 |
| Bank of Norway headquarters, Norway | 440 |
| Furka base tunnel, Switzerland | 300 |
| Verrazano Narrow bridge, USA | 280 |
| Boston's Big Dig artery/tunnel project, USA | 220 |
| Denver international airport, USA | 200 |
| Panama canal, Panama | 200 |
| Minneapolis Hiawatha light rail line, USA | 190 |
| Humber bridge, UK | 180 |
| Dublin Port tunnel, Ireland | 160 |
| Montreal metro Laval extension, Canada | 160 |
| Copenhagen metro, Denmark | 150 |
| Boston-New York-Washington railway, USA | 130 |
| Great Belt rail tunnel, Denmark | 120 |
| London Limehouse road tunnel, UK | 110 |
| Brooklyn bridge, USA | 100 |
| Shinkansen Joetsu high-speed rail line, Japan | 100 |
| Channel tunnel, UK, France | 80 |
| Karlsruhe-Bretten light rail, Germany | 80 |
| London Jubilee Line extension, UK | 80 |
| Bangkok metro, Thailand | 70 |
| Mexico City metroline, Mexico | 60 |
| High-speed Rail Line South, The Netherlands | 60 |
| Great Belt east bridge, Denmark | 50 |

# Notes

[1] As a general rule of thumb, "megaprojects" are measured in billions of dollars, "major projects" in hundreds of millions, and "projects" in millions and tens of millions. Megaprojects are sometimes also called "major programs."

[2] The colleague is Dr. Patrick O'Connell, Practitioner Director of Major Program Management at Oxford University's Saïd Business School.

[3] "Uniqueness bias" is here defined as the tendency of planners and managers to see their projects as singular. This particular bias stems from the fact that new projects often use non-standard technologies and designs, leading managers to think their project is more different from other projects than it actually is. Uniqueness bias impedes managers' learning, because they think they have nothing to learn from other projects as their own project is unique. This lack of learning may explain why managers who see their projects as unique perform significantly worse than other managers (Budzier and Flyvbjerg 2013). Project managers who think their project is unique are therefore a liability for their project and organization. For megaprojects this would be a mega-liability.

[4] Quoted from "Under Water Over Budget," *The Economist*, 7 October 1989, 37–8.

[5] *The Economist* (March 10, 2012: 55) describes the near-certainty of large cost overruns and delays in transportation infrastructure projects as "the iron law of infrastructure projects." Our data show the iron law is not limited to infrastructure; it applies to megaprojects in general and covers benefit shortfalls in addition to cost overruns and delays.

[6] The author owes the term "break-fix model" to Dr. Patrick O'Connell, Practitioner Director of Oxford University's BT Centre for Major Programme Management.

[7] Two versions of Hirschman's text exist (1967a, 1967b). The version of the text referenced here is the one published in *Development Projects Observed* (Hirschman 1967a), which is the original text. The differences between the two texts are minor and are mainly due to the editing of Irving Kristol, editor of *The Public Interest* at the time of publication (Adelman 2013: 405).

[8] The Sarbanes-Oxley Act of 2002 pioneered this area in the US, but many other countries have since followed suit with similar legislation. Section 802[a] (18 U.S.C. § 1519) of the original act states that whoever knowingly alters, destroys, mutilates, conceals, covers up, falsifies, or makes a false entry in any record, document, or tangible object with the intent to impede, obstruct, or influence the investigation or proper administration of any matter within the jurisdiction of any department or agency of the United States or any case filed under title 11, or in relation to or contemplation of any such matter or case, shall be fined, imprisoned not more than 20 years, or both.

[9] For full disclosure: The author was involved in the planning, start up, and delivery of the UK Major Projects Leadership Academy.

[10] See Flyvbjerg et al. (2012) regarding the use of tension points for triggering change in policy and practice, including for megaprojects.